\RequirePackage[mathlines]{lineno}
\documentclass[prl,twocolumn,showpacs,amsmath,amssymb]{revtex4-1}
\usepackage{graphicx}
\usepackage{epsfig,graphics,subfigure,psfrag,amsmath,amssymb}
\usepackage{dcolumn}
\usepackage{bm}
\usepackage{overpic}
\usepackage{tabularx}
\usepackage{color}
\usepackage{multirow}
\usepackage{colortbl}
\usepackage{epstopdf}
\usepackage{appendix}
\usepackage{rotating}
\usepackage{sidecap}
\usepackage{ulem}
\usepackage[colorlinks,linkcolor=blue,anchorcolor=blue,citecolor=blue]{hyperref}

\def\foEta{0.4576\pm0.0054_{\rm stat}\pm0.0045_{\rm syst}}
\def\foEtap{0.490\pm0.050_{\rm stat}\pm0.011_{\rm syst}}
\def\brRatio{0.355\pm0.033_{\rm stat}\pm0.015_{\rm syst}}
\def\mixing{(40.1\pm2.1_{\rm stat}\pm0.7_{\rm syst})^\circ}
\def\Vcs{1.031\pm0.012_{\rm stat}\pm0.009_{\rm syst}\pm0.079_{\rm theo}}
\def\Vcsp{0.917\pm0.094_{\rm stat}\pm0.021_{\rm syst}\pm0.155_{\rm theo}}

\begin{document}
\title{{\bf Measurement of the Dynamics of the Decays ${ D_s^+ \rightarrow \eta^{(\prime)}  e^{+} \nu_e}$}}


\author{
\small
M.~Ablikim$^{1}$, M.~N.~Achasov$^{10,d}$, S.~Ahmed$^{15}$, M.~Albrecht$^{4}$, M.~Alekseev$^{55A,55C}$, A.~Amoroso$^{55A,55C}$, F.~F.~An$^{1}$, Q.~An$^{52,42}$, J.~Z.~Bai$^{1}$, Y.~Bai$^{41}$
% ...
\\
\vspace{0.2cm}
(BESIII Collaboration)\\
\vspace{0.2cm} {\it
$^{1}$ Institute of High Energy Physics, Beijing 100049, People's Republic of China\\
% ...
$^{4}$ Bochum Ruhr-University, D-44780 Bochum, Germany\\
% ...
$^{10}$ G.I. Budker Institute of Nuclear Physics SB RAS (BINP), Novosibirsk 630090, Russia\\
% ...
$^{15}$ Helmholtz Institute Mainz, Johann-Joachim-Becher-Weg 45, D-55099 Mainz, Germany\\
% ...
$^{41}$ Southeast University, Nanjing 211100, People's Republic of China\\
$^{42}$ State Key Laboratory of Particle Detection and Electronics, Beijing 100049, Hefei 230026, People's Republic of China\\
% ...
$^{52}$ University of Science and Technology of China, Hefei 230026, People's Republic of China\\
% ...
$^{55}$ (A)University of Turin, I-10125, Turin, Italy; (B)University of Eastern Piedmont, I-15121, Alessandria, Italy; (C)INFN, I-10125, Turin, Italy\\
% ...
$^{d}$ Also at the Novosibirsk State University, Novosibirsk, 630090, Russia\\
% ...
}
}

\date{\today}

\begin{abstract}
  Using $e^+e^-$ annihilation data corresponding to an integrated luminosity of 3.19\,fb$^{-1}$
  collected at a center-of-mass energy of 4.178\,GeV with the BESIII detector, we measure the
  absolute branching fractions $\mathcal{B}_{D_s^+ \rightarrow \eta e^{+} \nu_e }$ = $(2.323\pm0.063_{\rm stat}\pm0.063_{\rm syst})\%$ and
  $\mathcal{B}_{D_s^+ \rightarrow \eta^{\prime} e^{+} \nu_e}$ = $(0.824\pm0.073_{\rm stat}\pm0.027_{\rm syst})\%$ via a tagged analysis
  technique, where one $D_s$ is fully reconstructed in a hadronic mode. Combining these measurements
  with previous BESIII measurements of $\mathcal{B}_{D^+\to\eta^{(\prime)} e^{+} \nu_e}$, the
  $\eta-\eta^\prime$ mixing angle in the quark flavour basis is determined to be
  $\phi_{\rm P} = (40.1\pm2.1_{\rm stat}\pm0.7_{\rm syst})^\circ$. From the first measurements of the dynamics of
  $D^+_s\to \eta^{(\prime)}e^+\nu_e$ decays, the products of the hadronic form factors
  $f_+^{\eta^{(\prime)}}(0)$ and the Cabibbo-Kobayashi-Maskawa matrix element $|V_{cs}|$ are determined with different
  form factor parametrizations. For the two-parameter series expansion, the results are
  $f^{\eta}_+(0)|V_{cs}| = 0.4455\pm0.0053_{\rm stat}\pm0.0044_{\rm syst}$ and $f^{\eta^{\prime}}_+(0)|V_{cs}| = 0.477\pm0.049_{\rm stat}\pm0.011_{\rm syst}$.
\end{abstract}

\maketitle

\oddsidemargin  -0.2cm
\evensidemargin -0.2cm

Exclusive $D$ semi-leptonic~(SL) decays provide a powerful way to extract the weak and strong
interaction couplings of quarks due to simple theoretical treatment~\cite{Riggio:2017zwh,Zhang:2018jtm,Fang:2014sqa}.  
In the Standard Model, the rate of $D^+_s\to \eta e^+\nu_e$ and $D^+_s\to \eta^{\prime}e^+\nu_e$ 
depends not only on $V_{cs}$, an element of the Cabibbo-Kobayashi-Maskawa (CKM) matrix 
describing weak transitions between the charm and strange quarks, but also on the dynamics of 
strong interaction, parameterized by the form factor~(FF) $f_+^{\eta^{(\prime)}}(q^2)$, where $q$ is the 
momentum transfer to the $e^+\nu_e$ system. Unlike the final-state hadrons $K$ and $\pi$, the 
mesons $\eta^{(\prime)}$ are especially intriguing because the spectator quark plays an important 
role in forming the final state.  This gives access to the singlet-octet mixing of $\eta-\eta^{\prime}$-gluon~\cite{Christ:2010dd,Dudek:2011tt}, 
whose mixing parameter can be determined from the SL decays, and, consequently, gives a deeper
understanding of non-perturbative QCD confinement.

Recently, the FFs $f^{\eta^{(\prime)}}_+(0)$ were calculated using lattice quantum
chromodynamics~(LQCD)~\cite{Bali:2014pva} and QCD light-cone sum rules
(LCSR)~\cite{Offen:2013nma,Duplancic:2015zna} by assuming particular admixtures of quarks and gluons
\cite{DiDonato:2011kr,Ambrosino:2006gk,Aaij:2014jna} for $\eta$ and $\eta^\prime$ mesons.  As
information concerning the gluon content in the $\eta^\prime$ remains inconclusive, large
uncertainties may be involved.  Measurements of $f^{\eta^{(\prime)}}_+(0)$ are crucial to calibrate
these theoretical calculations.  Once the predicted $f^{\eta^{(\prime)}}_+(0)$ pass these
experimental tests, they will help determine $|V_{cs}|$, and, in return, help test the unitarity of
the CKM quark mixing matrix.  Additionally, measurements of the branching fractions (BFs) of
$D_s^+\to \eta^{(\prime)}e^+\nu_e$ can shed light on $\eta-\eta^\prime$-gluon mixing.  The
$\eta-\eta^\prime$ mixing angle in the quark flavour basis, $\phi_{\rm P}$, can be related to the
BFs of the $D$ and $D_s$ via
$\cot^4\phi_{\rm P}=\frac{\Gamma_{D^+_s\to\eta^\prime e^+\nu_e}/\Gamma_{D^+_s\to\eta
    e^+\nu_e}}{\Gamma_{D^+\to\eta^\prime e^+\nu_e}/\Gamma_{D^+\to\eta e^+\nu_e}}$, in which a
possible gluon component cancels~\cite{DiDonato:2011kr}.  Determination of $\phi_{\rm P}$ gives a
complementary constraint on the role of gluonium in the $\eta^\prime$, thus helping to improve our
understanding of nonperturbative QCD dynamics and benefiting theoretical calculations of $D$ and $B$
decays involving the $\eta^{(\prime)}$.

Previous measurements of the BFs of $D^+_s\to\eta^{(\prime)}e^+\nu_e$ were made by CLEO
\cite{Brandenburg:1995qq,Yelton:2009aa,Hietala:2015jqa} and BESIII~\cite{Ablikim:2016rqq}, but these
measurements include large uncertainties.  This Letter reports improved measurements of the BFs and
the first experimental studies of the dynamics of $D^+_s\to \eta^{(\prime)} e^+\nu_e$
\cite{ref:ChargeConjugation}.  Based on these, the first measurements of $f_+^{\eta^{(\prime)}}(0)$
are made, and measurements of $|V_{cs}|$ and $\phi_{\rm P}$ are presented.

This analysis is performed using $e^+e^-$ collision data corresponding to an integrated luminosity
of 3.19\,fb$^{-1}$ taken at a center-of-mass energy $E_{\rm CM}=4.178$\,GeV with the BESIII
detector.  A description of the design and performance of the BESIII detector can be found in
Ref.~\cite{Ablikim2010345}. For the data used in this Letter, the end cap time-of-flight system was
upgraded with multi-gap resistive plate chambers with a time resolution of
60\,ps~\cite{Lxin,Gyingxiao}.  Monte Carlo (MC) simulated events are generated with a
{\sc{geant4}}-based~\cite{Agostinelli:2002hh} detector simulation software package, which includes
the geometric description and a simulation of the response of the detector. An inclusive MC sample
with equivalent luminosity 35 times that of data is produced at $E_{\rm CM} = 4.178$~GeV. It
includes open charm processes, initial state radiation (ISR) production of $\psi(3770)$,
$\psi(3686)$ and $J/\psi$, $q\bar q\,(q=u,\,d,\,s$) continuum processes, along with Bhabha
scattering, $\mu^+\mu^-$, $\tau^+\tau^-$ and $\gamma\gamma$ events.  The open charm processes are
generated using {\sc{conexc}}~\cite{Ping:2013jka}. The effects of ISR and final state radiation
(FSR) are considered. The known particle decays are generated with the BFs taken from the Particle
Data Group (PDG)~\cite{Tanabashi:2018oca} by {\sc{evtgen}}~\cite{ref:evtgen}, and the other modes
are generated using {\sc{lundcharm}} \cite{ref:lundcharm}. The SL decays
$D^+_s\to \eta^{(\prime)}e^+\nu_e$ are simulated with the modified pole
model~\cite{Chikilev:1999zn}.

At $E_{\rm CM}=4.178$~GeV, $D_s^+$ mesons are produced mainly from the processes
$e^+e^-\to D^+_sD_s^{*-} + c.c.\to D^+_s\gamma(\pi^0)D_s^-$. We first fully reconstruct one $D_s^-$
in one of several hadronic decay modes [called the single-tag (ST) $D^-_s$].  We then examine the SL
decays of the $D^+_s$ and the $\gamma(\pi^0)$ from the $D_s^{*}$ [called double-tag~(DT) $D_s^+$].
The BF of the SL decay is determined by

\begin{equation}
{\mathcal B}_{\rm SL} = N_{\rm DT}^{\rm tot}/(N^{\rm tot}_{\rm ST}\times\epsilon_{\gamma(\pi^0)\rm SL}),
\end{equation}
where $N^{\rm tot}_{\rm ST}$ and $N^{\rm tot}_{\rm DT}$ are the ST and DT yields in data,
$\epsilon_{\gamma(\pi^0)\rm SL}$ is the efficiency of finding $\gamma(\pi^0)\eta^{(\prime)}e^+\nu_e$
determined by
$\sum_k\frac{N^k_{\rm ST}}{N^{\rm tot}_{\rm ST}}\frac{\epsilon^k_{\rm DT}}{\epsilon^k_{\rm ST}}$,
where $\epsilon^k_{\rm ST}$ and $\epsilon^k_{\rm DT}$ are the efficiencies of selecting ST and DT
candidates in the $k$-th tag mode, and estimated by analyzing the inclusive MC sample and the independent signal MC events of various DT modes, respectively.

\begin{figure}
  \centering
  \includegraphics[width=8.5cm,height=8.5cm]{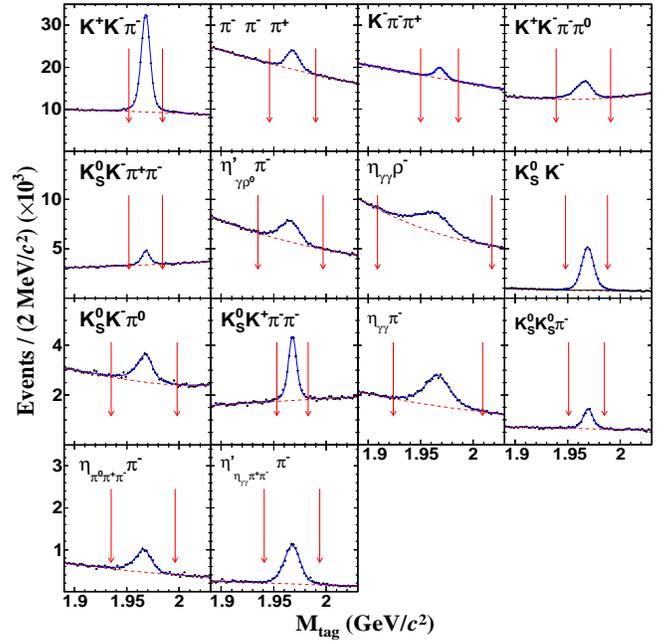}
  \caption{\label{fig:ST}\small Spectra of $M_{\rm tag}$ of the ST
    candidates. Dots with error bars are data. Blue solid curves are
    the fit results. Dashed curves are the fitted backgrounds. The
    black solid curve in the $K_S^0K^-$ mode is $D^-\to K_S^0\pi^-$
    background. Pairs of arrows denote the $D_s^-$ signal regions  within $\pm 3\sigma$ around the nominal $D^-_s$ mass~\cite{Tanabashi:2018oca}.}
\end{figure}

The ST $D^-_s$ candidates are reconstructed using fourteen hadronic decay modes as shown in
Fig.~\ref{fig:ST}. The selection criteria for charged tracks and $K^0_S$, and the particle
identification (PID) requirements for $\pi^\pm$ and $K^\pm$, are the same as those used in
Ref.~\cite{Ablikim:2018jun}. Positron PID is performed by using the specific ionization energy loss
in the main drift chamber, the time of flight, and the energy deposited in the electromagnetic
calorimeter~(EMC). Confidence levels for the pion, kaon and positron hypotheses (${\mathcal L}_\pi$,
${\mathcal L}_K$ and ${\mathcal L}_e$) are formed.  Positron candidates must satisfy
${\mathcal L}_e>0.001$ and ${\mathcal L}_e/({\mathcal L}_e+{\mathcal L}_\pi+{\mathcal
  L}_K)>0.8$. The energy loss of the positron due to bremsstrahlung is partially recovered by adding
the energies of the EMC showers that are within $10^\circ$ of the positron direction and not matched
to other particles (FSR recovery).

Photon candidates are selected from the EMC showers that begin within 700 ns of the event start time
and have an energy greater than 25 (50) MeV in the barrel (endcap) region of the
EMC~\cite{Ablikim2010345}. Candidates of $\pi^0$ or $\eta_{\gamma\gamma}$ are formed by photon pairs
with an invariant mass in the range (0.115,~0.150) or (0.50,~0.57)~GeV/$c^2$. To improve the
momentum resolution, the $\gamma\gamma$ invariant mass is constrained to the $\pi^0$ or $\eta$
nominal mass~\cite{Tanabashi:2018oca} via a kinematic fit.  Candidates of $\eta_{\pi^0\pi^+\pi^-}$,
$\eta'_{\eta_{\gamma\gamma}\pi^+\pi^-}$, $\eta'_{\gamma\rho^0}$, $\rho^0$, and $\rho^-$ are formed
from $\pi^+\pi^-\pi^0$, $\eta_{\gamma\gamma}\pi^+\pi^-$,$\gamma\rho^0_{\pi^+\pi^-}$, $\pi^+\pi^-$
and $\pi^-\pi^0$ combinations whose invariant masses fall in the ranges $(0.53,\,0.57)$,
$(0.946,\,0.970)$, $(0.940,0.976)$, $(0.57,0.97)$, and $(0.57,\,0.97)$~GeV/$c^2$, respectively.

To remove soft pions originating from $D^*$ transitions, the momenta of pions from the ST $D_s^-$
are required to be larger than 0.1~GeV/$c$.  For the tag modes $D_s^-\to\pi^+\pi^-\pi^-$ and
$K^-\pi^+\pi^-$, the contributions of $D^-_s\to K^0_S\pi^-$ and $K^0_SK^-$ are removed by requiring
$M_{\pi^+\pi^-}$ outside $\pm0.03$\,GeV/$c^2$ around the $K^0_S$ nominal mass~\cite{Tanabashi:2018oca}.

The ST $D^-_s$ mesons are identified by the beam constrained mass
$M_{\rm BC}\equiv\sqrt{(E_{\rm c.m.}/2)^2 - |\vec{p}_{D_s^-}|^2}$ and the $D_s^-$ recoil mass
$M_{\rm rec} \equiv \sqrt{(E_{\rm c.m.} - \sqrt{|\vec{p}_{D_s^-}|^2 + M_{D_s^-}^2})^2 -
  |\vec{p}_{D_s^-}|^2}$, where $\vec{p}_{D^-_s}$ is the 3-momentum of the ST candidate and
$M_{D^-_s}$ is the nominal $D^-_s$ mass \cite{Tanabashi:2018oca}. Non-$D^+_s D^{*-}_s$ events are
suppressed by requiring $M_{\rm BC} \in(2.010,\,2.073)$~GeV/$c^2$. In each event, only the candidate
with $M_{\rm rec}$ closest to the nominal $D_s^{*+}$ mass~\cite{Tanabashi:2018oca} is chosen.  The
ST yield is determined by fits to the $M_{\rm tag}$ spectra for each of the 14 tag modes shown in
Fig.~\ref{fig:ST}, where $M_{\rm tag}$ is the invariant mass of the ST candidate.  Signals and the
$D^-\to K^0_S\pi^-$ peaking background in the $D^-_s\to K^0_SK^-$ mode are described by MC-simulated
shapes. The nonpeaking background is modeled by a second- or third-order Chebychev polynomial. To
account for the resolution difference between data and MC simulation, the MC simulated shape(s) is
convolved with a Gaussian for each tag mode. The reliability of the fitted nonpeaking background
has been verified using the inclusive MC sample.  Events in the signal regions, denoted by the
boundaries in each subfigure of Fig.~\ref{fig:ST}, are kept for further analysis. The total ST
yield is $N^{\rm tot}_{\rm ST}=395142\pm1923$.

Once the $D_s^-$ tag has been found, the photon or $\pi^0$ from the $D_s^{*+}$ transition is
selected. We define the energy difference
$\Delta E \equiv E_{\rm CM} - E_{\rm tag} - E_{\gamma(\pi^0)+D^-_s}^{\rm rec} - E_{\gamma(\pi^0)}$,
where
$E_{\gamma(\pi^0)+D^-_s}^{\rm rec} \equiv \sqrt{|-\vec{p}_{\gamma(\pi^0)}-\vec{p}_{\rm tag}|^2 +
  M^2_{D_s^+}}$, $E_i$ and $\vec{p}_i$ [$i = \gamma(\pi^0)$ or tag] are the energy and momentum of
$\gamma(\pi^0)$ or $D_s^-$ tag, respectively. All unused $\gamma$ or $\pi^0$ candidates are looped
over and that with the minimum $|\Delta E|$ is chosen. Candidates with
$\Delta E\in(-0.04,\,0.04)$\,GeV are accepted. The signal candidates are examined by the kinematic
variable
$ {\rm MM}^2 \equiv (E_{\rm CM} - E_{\rm tag} - E_{\gamma(\pi^0)} - E_{\eta^{(\prime)}} - E_e)^2 -
|-\vec{p}_{\rm tag} - \vec{p}_{\gamma(\pi^0)} - \vec{p}_{\eta^{(\prime)}} - \vec{p}_e|^2,$ where
$E_i$ and $\vec{p}_i$ ($i = e$ or $\eta^{(\prime)}$) are the energy and momentum of $e^+$ or
$\eta^{(\prime)}$. To suppress backgrounds from $D^+_s$ hadronic decays, the maximum energy of the
unused showers ($E_{\rm \gamma~extra }^{\rm max}$) must be less than 0.3\,GeV and events with
additional charged tracks ($N^{\rm extra}_{\rm char}$) are removed. We require
$M_{\eta^\prime e^+}< 1.9$\,GeV/$c^2$ for $D_s^+\to\eta^\prime e^+\nu_e$ and
$\cos\theta_{\rm hel}\in(-0.85,\,0.85)$ for $D^+_s\to \eta^\prime_{\gamma\rho^0}e^+\nu_e$ to further
suppress the $D_s^+\to\eta^\prime \pi^+$ and $D_s^+\to\phi e^+\nu_e$ backgrounds, where
$\theta_{\rm hel}$ is the helicity angle between the momentum directions of the $\pi^+$ and the $\eta^\prime$
in the $\rho^0$ rest frame.

\begin{figure}[htp]
    \centering
 \includegraphics[width=8.5cm]{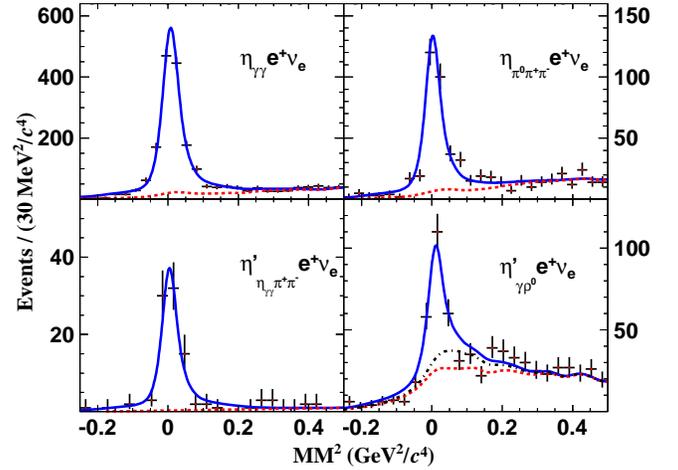}\\
   \caption{Distributions of MM$^2$ of the SL candidates. Dots with error bars are data. Solid curves are the best fits. Dotted curves are the fitted non-peaking backgrounds. The dash-dotted curve is the peaking background due to $D^+_s\to \phi e^+\nu_e$. \label{fig:signal_yields_fromdata}}
\end{figure}

Figure~\ref{fig:signal_yields_fromdata} shows the MM$^2$ distribution after all selection criteria
have been applied. The signal yields are determined from a simultaneous unbinned maximum likelihood
fit to these spectra, where $\mathcal{B}_{D_s^+\to\eta^{(\prime)} e^+\nu_e}$ measured using two
different $\eta^{(\prime)}$ subdecays are constrained to be the same after considering the different
efficiencies and subdecay BFs.  The signal and background components in the fit are described by
shapes derived from MC simulation.  For the decay $D_s^+ \to \eta^\prime_{\gamma\rho^0} e^+ \nu_e$,
some peaking background from $D_s^+ \to \phi e^+ \nu_e$ still remains.  This background is modeled
by a separate component in the fit; its size and shape are fixed based on MC simulation.

Table~\ref{table:br} summarizes the efficiencies for finding SL decays, the observed signal yields,
and the obtained BFs.

   \begin{table}
   \centering
    \caption{\small Efficiencies ($\epsilon_{\rm \gamma(\pi^0){\rm SL}}$), signal yields ($N_{\rm DT}^{\rm tot}$), and the obtained BFs. Uncertainties on the least significant digits are shown in parentheses, where the first (second) uncertainties are statistical (systematic). The efficiencies do not include the BFs of $\eta^{(\prime)}$ subdecays. \label{table:br}}
           \begin{tabular}[t]{ccccccccc}\hline\hline
        Decay  & $\eta^{(\prime)}$ decay  &$\epsilon_{\rm \gamma(\pi^0){\rm SL}}$~(\%)  &$N_{\rm DT}^{\rm tot}$ &$\mathcal{B}_{\rm SL}$~(\%)\\\hline
        \multirow{2}{*}{$\eta e^+\nu_e$}  & $\gamma\gamma$   & 41.11(27)  & \multirow{2}{1.2cm}{1834(47)} & \multirow{2}{2.2cm}{2.323(63)(63)}\\
                                                         & $\pi^0\pi^+\pi^-$        & 16.06(31) &                                & \\\hline
        \multirow{2}{*}{$\eta^\prime e^+\nu_e$}&      $\eta\pi^+\pi^-$       & 14.07(10) & \multirow{2}{1.2cm}{261(22)}  & \multirow{2}{2.2cm}{0.824(73)(27)}\\
                                                         & $\gamma\rho^0$       & 18.98(10)  &                                & \\\hline\hline
        \end{tabular}
\end{table}


With the DT method, the BF measurements are insensitive to the ST selection. The following relative
systematic uncertainties in the BF measurements are assigned.  The uncertainty in the ST yield is
estimated to be 0.6\% by alternative fits to the $M_{\rm tag}$ spectra with different signal shapes,
background parameters, and fit ranges. The uncertainties in the tracking or PID efficiencies are
assigned as 0.5\% per $\pi^\pm$ by studying $e^+e^-\to K^+K^-\pi^+\pi^-$, and 0.5\% per $e^+$ by
radiative Bhabha process, respectively.  The uncertainties of the $E_{\rm \gamma~extra }^{\rm max}$
and $N_{\rm char}^{\rm extra}$ requirements are estimated to be 0.5\% and 0.9\% by analyzing DT
hadronic events.  The uncertainties of the $\Delta E$ requirement, FSR recovery and
$\theta_{\rm hel}$ requirement are estimated with and without each requirement, and the BF changes
are 0.8\%, 0.8\%, and 0.1\%, respectively, which are taken as the individual uncertainties. The
uncertainties of the selection of neutral particles are assigned as 1.0\% per photon by studying
$J/\psi\to \pi^+\pi^-\pi^0$~\cite{Ablikim:2011kv} and 1.0\% per $\pi^0$ or $\eta$ by studying
$e^+e^-\to K^+K^-\pi^+\pi^-\pi^0$.  The uncertainty due to the signal model is estimated to be 0.5\% by
comparing the DT efficiencies before and after re-weighting the $q^2$ distribution of the signal MC
events to data.  The uncertainty of the ${\rm MM}^2$ fit is assigned as 0.9\%, 1.3\%, 1.2\% and
1.2\% for $D^+_s\to \eta_{\gamma\gamma}e^+\nu_e$, $\eta_{\pi^0\pi^+\pi^-}e^+\nu_e$,
$\eta^\prime_{\eta\pi^+\pi^-}e^+\nu_e$ and $\eta^\prime_{\gamma\rho^0}e^+\nu_e$ (the same sequence
later), respectively, by repeating fits with different fit ranges and different signal and
background shapes.
The ST efficiencies may be different due to the different
multiplicities in the tag environments, leading to incomplete cancelation
of the systematic uncertainties associated with the ST selection.
The associated uncertainty is assigned as 0.4\%, 0.3\%, 0.3\%, 0.3\%,
from studies of the
efficiency differences for tracking and PID of $K^\pm$ and
$\pi^\pm$ as well as the selection of neutral particles between
data and MC simulation in different environments.
The uncertainty due to the $M_{\eta^\prime e^+}$
requirement is found to be negligible.  The uncertainty due to peaking background is assigned to be
1.4\% by varying its size by $\pm1 \sigma$ of the corresponding BF. The
uncertainties due to the quoted BFs, 0.9\%, 1.4\%, 1.8\% and 1.9\%~of $\eta^{(\prime)}$
decays~\cite{Tanabashi:2018oca} are also considered. For each decay, the total systematic
uncertainty is determined to be 2.7\%, 3.3\%, 3.4\% and 4.0\% by adding all these uncertainties in
quadrature.


With the BFs measured in this work, we determine the BF ratio
${\mathcal R}^{D^+_s}_{\eta^\prime/\eta}={\mathcal B}_{D^+_s\to\eta e^+\nu_e}/{\mathcal
  B}_{D^+_s\to\eta^\prime e^+\nu_e} = \brRatio$, where the systematic uncertainties on the ST yield
and due to the photon from $D_s^{*+}$, FSR recovery, tracking and PID of $e^+$ cancel. Using these
BFs and $\mathcal{B}_{D^+\to\eta^{(\prime)} e^{+} \nu_e}$ reported in
Ref.~\cite{Zhangyu:Dptoetaenu}, we determine the $\eta-\eta^\prime$ mixing angle to be
$\phi_P= \mixing$. This result is consistent with previous measurements using
$D\to \eta^{(\prime)} e^+\nu_e$ decays~\cite{DiDonato:2011kr} and $\psi\to \gamma \eta^{(\prime)}$
decays~\cite{Ambrosino:2006gk} within uncertainties.


To study the $D^+_s\to\eta^{(\prime)} e^+\nu_e$ dynamics, the candidate events are divided into
various $q^2$ intervals. The measured partial decay width $\Delta\Gamma^i_{\rm msr}$ in the $i$th
$q^2$ interval is determined by
$\Delta \Gamma^i_{\rm msr} \equiv \int_i\frac{d\Gamma}{dq^2}dq^2 = \frac{N_{\rm pro}^i}{\tau_{D_s^+}
  \times N^{\rm tot}_{\rm ST}}$, where $\tau_{D_s^+}$ is the lifetime of the $D_s^+$
meson~\cite{Tanabashi:2018oca,Aaij:2017vqj}, and $N^i_{\rm pro}$ is the DT yield produced in the
$i$th $q^2$ interval, calculated by
$N_{\rm pro}^{i} = \sum^{m}_{j}(\epsilon^{-1})_{ij} N_{\rm obs}^{j}$. Here $m$ is the number of
$q^2$ intervals, $N^j_{\rm obs}$ is the observed DT yield obtained from similar fits to the MM$^2$
distribution as described previously, and $\epsilon_{ij}$ is the efficiency matrix determined from
signal MC events and is given by
 $\epsilon_{ij} = \sum_k\left[(1/N_{\rm ST}^{\rm tot})\times(N^{ij}_{\rm rec}/N^j_{\rm gen})_k
\times (N^k_{\rm ST}/\epsilon^k_{\rm ST})\right]$, where $N^{ij}_{\rm rec}$ is the DT yield generated
in the $j$th $q^2$ interval and reconstructed in the $i$th $q^2$ interval, $N^j_{\rm gen}$ is the total signal 
yield generated in the $j$th $q^2$ interval, and $k$ sums over all tag modes. See Tables 1 and 2 of Ref. [30] for details 
about the range, $N^i_{\rm obs}$, $N^i_{\rm prd}$, and $\Delta\Gamma^i_{\rm msr}$ of each $q^2$ interval for $D^+_s\to \eta e^+\nu_e$ and $D^+_s\to \eta^\prime e^+\nu_e$, respectively. 

In theory, the differential decay width can be expressed 
\begin{equation}
    \frac{d\Gamma(D_s^+\rightarrow\eta^{(\prime)}e^+\nu)}{dq^2} = \frac{G_F^2|V_{cs}|^2}{24\pi^3}|f_+^{\eta^{(\prime)}}(q^2)|^2|p_{\eta^{(\prime)}}|^3,
\end{equation}  
where $|p_{\eta^{(\prime)}}|$ is the magnitude of the meson 3-momentum in the $D_s^+$ rest frame and $G_F$ is the Fermi constant. 
In the modified pole model~\cite{Becher:2005bg},

\begin{equation}
    f_+(q^2) = \frac{f_+(0)}{(1-\frac{q^2}{M^2_{\rm pole}})(1-\alpha\frac{q^2}{M^2_{\rm pole}})},
    \label{eq:mod_pole}
\end{equation}
where $M_{\rm pole}$ is fixed to $M_{D^{*+}_s}$ and $\alpha$ is a free parameter. Setting $\alpha=0$~and leaving $M_{\rm pole}$ free, it is the simple pole model~\cite{Becirevic:1999kt}. In the two-parameter (2 Par.) series expansion~\cite{Becher:2005bg}

\begin{equation}
    f_+(q^2) = \frac{1}{A(q^2)}\frac{f_+(0)A(0)}{1+B(0)}(1 + B(q^2)).
    \label{eq:Series_model}
\end{equation}
Here, $A(q^2)=P(q^2)\Phi(q^2,t_0)$, $B(q^2)=r_1(t_0)z(q^2,t_0)$, $t_0 = t_+(1-\sqrt{1-t_-/t_+})$, $t_{\pm} = (M_{D_s^+}\pm M_{\eta})$, and $r_k$ is a free parameter. The functions $P(q^2)$, $\Phi(q^2,\,t_0)$, and $z(q^2,\,t_0)$ are defined following Ref.~\cite{Becher:2005bg}.

For each SL decay, the product $f_+(0)|V_{cs}|$ and one other parameter, $M_{\rm pole}$, $\alpha$, or $r_1$, are
determined by constructing and minimizing

\begin{equation}
    \chi^2 = \underset{ij=1}{\sum^{m}}(\Delta \Gamma^i_{\rm msr} - \Delta \Gamma^i_{\rm exp})
    C_{ij}^{-1}(\Delta \Gamma^j_{\rm msr} - \Delta \Gamma^j_{\rm exp}),
    \label{eq:chi_square}
\end{equation}
with $\Delta\Gamma^i_{\rm msr}$ and the theoretically expected value $\Delta\Gamma^i_{\rm exp}$,
where $C_{ij} = C_{ij}^{\rm stat}+C_{ij}^{\rm syst}$ is the covariance matrix of
$\Delta\Gamma^i_{\rm msr}$ among $q^2$ intervals, as shown in Tables 3 and 4 in
Ref.~\cite{Supplement}. For each $\eta^{(\prime)}$ subdecay, the statistical covariance matrix is
constructed with the statistical uncertainty in each $q^2$ interval ($\sigma(N^\alpha_{\rm obs})$)
as
$C_{ij}^{\rm stat} = (\frac{1}{\tau_{D_s^+}N_{\rm ST}^{\rm
    tot}})^2\sum_{\alpha}\epsilon_{i\alpha}^{-1}\epsilon_{j\alpha}^{-1}[\sigma(N^{\alpha}_{\rm
  obs})]^2$.  The systematic covariance matrix is obtained by summing all the covariance matrices
for all systematic uncertainties, which are all constructed with the systematic uncertainty in each
$q^2$ interval ($\delta(\Delta\Gamma^i_{\rm msr})$) as
$C_{ij}^{\rm syst} = \delta(\Delta \Gamma^i_{\rm msr})\delta(\Delta \Gamma^j_{\rm msr})$.  Here, an
additional systematic uncertainty in $\tau_{D^+_s}$ (0.8\%)~\cite{Tanabashi:2018oca,Aaij:2017vqj} is
involved besides those in the BF measurements.

\begin{figure}
\centering
     \mbox{
    \begin{overpic}[width=8.8cm,angle=0]{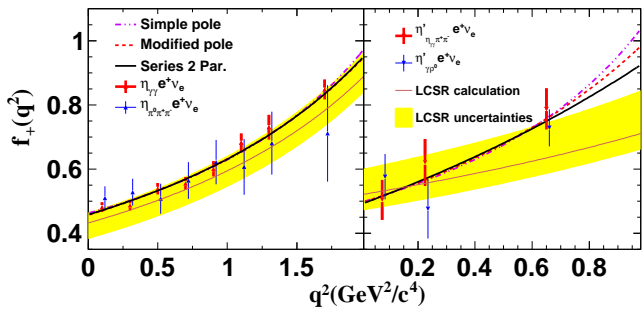}
    \put(35,30){{\large  }}
    \end{overpic}
    }
\caption{\small Projections of the fits to $\Delta\Gamma^i_{\rm msr}$ of $D^+_s\to
  \eta^{(\prime)}e^+\nu_e$. Dots with error bars are data. The $\Delta\Gamma^i_{\rm msr}$s measured with
  the two $\eta^{(\prime)}$ decay modes are offset horizontally for improved clarity. The curves
  show the best fits as described in text. Pink lines with yellow bands are the LCSR calculations with uncertainties~\cite{Offen:2013nma}. }
  \label{fig:combine_FF}
\end{figure}

The $\Delta\Gamma^i_{\rm msr}$ measured by the two $\eta^{(\prime)}$ subdecays are fitted
simultaneously, with results shown in Fig~\ref{fig:combine_FF}. In the fits, the
$\Delta\Gamma^i_{\rm msr}$ becomes a vector of length $2m$. Uncorrelated systematic uncertainties
are from tag bias, quoted BFs, $\eta$\,(and $\pi^0$) reconstruction, and FF
parametrization,~while other systematic uncertainties are fully correlated.
Table~\ref{tab:FF_combine} summarizes the fit results, where the obtained
$f_+^{\eta^{(\prime)}}(0)|V_{cs}|$ with different FF parameterizations are consistent with each
other.

Combining $|V_{cs}| = 0.97343 \pm 0.00015$ from the global fit in the SM~\cite{Tanabashi:2018oca}
with $f^{\eta^{(\prime)}}_+(0)|V_{cs}|$ extracted from the two-parameter series expansion, we
determine $f^{\eta}_+(0) = \foEta$ and $f^{\eta^{\prime}}_+(0)=\foEtap$. Table~\ref{table:FF}
compares the measured FFs with various theoretical calculations within uncertainties. When combining $f^{\eta}_+(0)$ and
$f^{\eta^{\prime}}_+(0)$ calculated from Ref.~\cite{Offen:2013nma}, we obtain $|V_{cs}|=\Vcs$ and
$\Vcsp$, respectively. These results agree with the measurements of $|V_{cs}|$ using
$D\to \bar K
\ell^+\nu_\ell$~\cite{Ablikim:2015ixa,Ablikim:2015qgt,Ablikim:2018evp,Besson:2009uv,Aubert:2007wg,Widhalm:2006wz}
and $D_s^+ \to \ell^+\nu_\ell$
decays~\cite{Ablikim:2016duz,Zupanc:2013byn,delAmoSanchez:2010jg,Onyisi:2009th,Naik:2009tk} within
uncertainties.

\begin{table*}
    \centering
    \caption{\small Results of the fits to $\Delta\Gamma^i_{\rm msr}$. Uncertainties on the least significant digits are shown in parentheses, where the first (second) uncertainties are statistical (systematic). $N_{\rm dof}$ is the number of degrees of freedom.\label{tab:FF_combine}}

    \begin{tabular}{|l|ccc|ccc|ccc|}\hline\hline
    Case &\multicolumn{3}{c|}{Simple pole}  &  \multicolumn{3}{c|}{Modified pole}  &  \multicolumn{3}{c|}{Series 2 Par.} \\
   &  $f_+^{\eta^{(\prime)}}(0)|V_{cs}|$           &  $M_{\rm pole}$        &   $\chi^2/{N_{\rm dof}}$& $f_+^{\eta^{(\prime)}}(0)|V_{cs}|$           &  $\alpha$              &   $\chi^2/{N_{\rm d.o.f}}$& $f_+^{\eta^{(\prime)}}(0)|V_{cs}|$           & $r_1$                  &   $\chi^2/{N_{\rm d.o.f}}$\\\hline
   ${ \eta e^{+} \nu_e}$&$0.4505(45)(31)$    &$3.759(84)(45)$    &12.2/14 &$0.4457(46)(34)$    &  $0.304(44)(22)$  &    11.4/14&$0.4465(51)(35)$    &$-2.25(23)(11)$      &11.5/14\\
   $\eta^{\prime} e^{+} \nu_e$&$0.483(42)(10) $   &$ 1.88(60)(08)$   &1.8/4 &$0.481(44)(10)$    &  $ 1.62(91)(13)$ &    1.8/4 &  $0.477(49)(11)$  & $-13.1(76)(10) $   &1.9/4\\\hline\hline
    \end{tabular}
\end{table*}

\begin{table*}
    \caption{Comparison of the measured $f_+^{\eta^{(\prime)}}(0)$ with the theoretical calculations. Errors on the least significant digits are shown in parentheses. For the LQCD model, the errors are statistical only, while $^{A(B)}$ assume $M_\pi=470(370)$~MeV.  \label{table:FF}}
  
        \begin{tabular}[t]{ccccccccccc}\hline\hline
          &CLFQM~\cite{Verma:2011yw}&CQM~\cite{Melikhov:2000yu}&CCQM~\cite{Soni:2018adu}&3PSR~\cite{Colangelo:2001cv}&LCSR~\cite{Azizi:2010zj}&LCSR~\cite{Offen:2013nma}& LQCD$^A$~\cite{Bali:2014pva}&LQCD$^B$~\cite{Bali:2014pva}   &        LCSR~\cite{Duplancic:2015zna}  & BESIII\\\hline
           $f^\eta_{\rm+}(0)$ &0.76&0.78&0.78(12)&0.50(4) & 0.45(15)&         0.432(33) & 0.564(11)                          & 0.542(13)                                 & 0.495(30)&  0.4576(70)\\
           $f^{\eta^\prime}_{\rm+}(0)$&-&0.78&0.73(11)&  -& 0.55(18)   &  0.520(88) & 0.437(18)            &   0.404(25)                             & 0.558(47) & 0.490(51)\\\hline\hline
        \end{tabular}
\end{table*}


In summary, by analyzing a data sample of 3.19 fb$^{-1}$ taken at $E_{\rm CM}=4.178$ GeV with the
BESIII detector, we measure the absolute BFs of $D^+_s\to \eta^{(\prime)} e^+\nu_e$ with a DT
method. The precision is improved by a factor of 2 compared to the world average values. Using these
BFs and ${\mathcal B}(D^+\to \eta^{(\prime)} e^+\nu_e)$ measured in our previous
work~\cite{Zhangyu:Dptoetaenu}, we determine the $\eta-\eta^\prime$ mixing angle $\phi_P$, which
provides complementary data to constrain the gluon component in the $\eta^\prime$ meson. From an
analysis of the dynamics in $D^+_s\to \eta^{(\prime)} e^+\nu_e$, the products of
$f_+^{\eta^{(\prime)}}(0)|V_{cs}|$ are determined for the first time. Furthermore, by taking $|V_{cs}|$
from a standard model fit (CKM{\sc fitter},~\cite{Tanabashi:2018oca}) as input, we determine the form
factor at zero momentum transfer $f^{\eta^{(\prime)}}_+(0)$ for
the first time. The obtained FFs provide important data to distinguish various theoretical
calculations~\cite{Bali:2014pva,Offen:2013nma,Duplancic:2015zna,Melikhov:2000yu,Soni:2018adu,Colangelo:2001cv,Azizi:2010zj}. Alternatively, we also determine
$|V_{cs}|$ with $D_s^+\to \eta^{(\prime)} e^+\nu_e$ decays for the first time, by taking values for
$f^{\eta^{(\prime)}}_+(0)$ calculated in theory. Our result on $|V_{cs}|$ together with those
measured by $D\to \bar K\ell^+\nu_\ell$ and $D_s^+\to \ell^+\nu_\ell$ are important to test the
unitarity of the CKM matrix.

\begin{acknowledgments}
The BESIII collaboration thanks the staff of BEPCII and the IHEP computing center for their strong support. This work is supported in part by National Key Basic Research Program of
China under Contract No. 2015CB856700; National Natural Science Foundation of China (NSFC) under Contracts Nos. 11335008, 11425524, 11625523, 11635010, 11735014;
the~Chinese~Academy of Sciences (CAS) Large-Scale Scientific Facility Program; the CAS Center for Excellence in Particle Physics (CCEPP); Joint Large-Scale Scientific Facility
Funds of the NSFC and CAS under Contracts Nos. U1632109, U1532257, U1532258, U1732263; CAS Key Research Program of Frontier Sciences under Contracts Nos. QYZDJ-SSW-SLH003,
QYZDJ-SSW-SLH040; 100 Talents Program of CAS; INPAC and Shanghai Key Laboratory for Particle Physics and Cosmology; German Research Foundation DFG under Contracts
Nos. Collaborative Research Center CRC 1044, FOR 2359; Istituto Nazionale di Fisica Nucleare, Italy; Koninklijke Nederlandse Akademie van Wetenschappen (KNAW) under Contract
No. 530-4CDP03; Ministry of Development of Turkey under Contract No. DPT2006K-120470; National Science and Technology fund; The Swedish Research Council; U. S. Department
of Energy under Contracts Nos. DE-FG02-05ER41374, DE-SC-0010118, DE-SC-0010504, DE-SC-0012069; University of Groningen (RuG) and the Helmholtzzentrum fuer
Schwerionenforschung GmbH (GSI), Darmstadt.
\end{acknowledgments}


\begin{thebibliography}{30}
\bibitem{ref:ChargeConjugation} Charge conjugated modes are implied throughout this Letter.
\end{thebibliography}
\end{document}